# Spontaneous Generation of Vortex Array Beams from a Thin-Slice Solid-State Laser with Wide-Aperture Laser-Diode Pumping

Kenju Otsuka[1] and Shu-Chun Chu[2]

[1]Department of Human and Information Science, Tokai University, 1117 Kitakaname, Hiratsuka, Kanagawa 259-1292, Japan

[2]Department of Physics, National Cheng Kung University, No. 1, University Road, Tainan City 701, Taiwan

**Abstract:** **We studied complex lasing pattern formations in a thin-slice solid-state laser with wide-aperture laser-diode end-pumping. Radial and rectangular vortex arrays were found to be formed in a controlled fashion with symmetric and asymmetric pump beam profiles, respectively. Most of these vortices exhibited single-frequency oscillations arising from a spontaneous process of transverse mode locking of nearly degenerate modes assisted by the laser nonlinearity. Single-frequency rectangular array beams consisting of a large number of vortices, e.g., closely packed 36 or 46 vortex pixels, were generated, originating from Ince-Gaussian modes excited by the asymmetric pumping.**

The formation of ordered transverse emission patterns such as hexagons [1], soliton pixels [2], and optical vortices crystals [3,4] is of fundamental interest and could also have applications in a variety of areas such as optical data storage, distribution, and processing that exploit the robustness of soliton and vortex fields [5] and optical manipulations of small particles and atoms in the featured intensity distribution [6].

The generation of global field patterns composed of bright or dark spots have been demonstrated in wide-aperture optical systems and lasers, which have high values of the Fresnel number and optical nonlinearity. The prototypical example for creating such complex patterns in lasers is the vertical cavity semiconductor laser diode (VCSEL), which possesses a large saturation-type of optical nonlinearity, i.e., light-intensity-dependent refractive index change. Theoretically, these nonlinear lasing pattern formations can be reproduced by exploring the rate equations [7] for wide-aperture lasers coupled to the nonlinear Maxwell's wave equation describing the transverse fields. Reducing these equations using paraxial field propagation and a steady-state assumption resulted in a complex Ginzburg Landau (CGL)-like equation with saturation-type nonlinearity. The formation of vortices crystals in a $Na_2$ laser and in a VCSEL was reported by Brambilla *et al*. [3] and Scheuer and Orenstein [4], respectively, and they interpreted these lasing patterns in terms of spontaneous transverse mode locking of nearly degenerated modes in quadrature assisted by the intrinsic nonlinearity of the laser.

On the other hand, Otsuka *et al.* have demonstrated complex lasing pattern formations in thin-slice solid-state lasers with laser-diode (LD) end-pumping and observed optical-billiard-like lasing patterns consisting of non-orthogonal transverse modes accompanied by modal-interference-induced modulations with the tightly focused strongly asymmetric LD pump beam, i.e., sheet-like pumping [8]. In this paper, we report on our systematic study of the effects of LD pump-beam shape and pump power on pattern formations in wide-aperture solid-state lasers, paying special attention to vortex array beam generation due to the inherent optical nonlinearity in detuned lasers.

The experimental setup is shown in Fig. 1(a). A nearly collimated beam from a laser diode operating at a wavelength of 808 nm passed through an objective lens with a numerical aperture N.A. = 0.25 and impinged on the $LiNdP_4O_{12}$ (LNP) laser. The active region of the LD we used in the present experiment was 1 µm thick and 100 µm wide. The end surfaces of the 7-mm-square, 0.3-mm-thick platelet LNP crystal were coated with dielectric mirrors $M_1$ (transmission at 808 nm > 95%; reflectance at 1048 nm = 99.8%) and $M_2$ (reflectance at 1048 nm = 99%). By changing the crystal position along the optical axis, z, we varied the pump-beam size and shape via the aberration of beam-focus. The pump-beam profile at the crystal was measured with a beam profiler. At the focal position, the focused beam shape was highly asymmetrical, i.e., sheet-like pumping. The measured beam sizes along the x and y axes ($w_x$, $w_y$) are shown in Fig. 1(b) as a function of crystal position, where z = 0 corresponds to the focal position. When the LNP crystal was shifted toward the LD, the pump beam aperture increased and its profile changed as depicted in Fig. 1(b). Far-field lasing patterns were observed with a PbS infrared viewer monitored on a TV, where lasing lateral distributions were preserved in propagation. The global oscillation spectrum measured with a multi-wavelength meter indicated single longitudinal-mode oscillations at a wavelength of 1048 nm. Linearly polarized operations along the x-axis, i.e., pseudo-orthorhombic c-axis of the LNP, were obtained. Detailed optical spectra were measured by a scanning Fabry-Perot interferometer (free spectral range: 2 GHz; frequency resolution: 6.6 MHz).

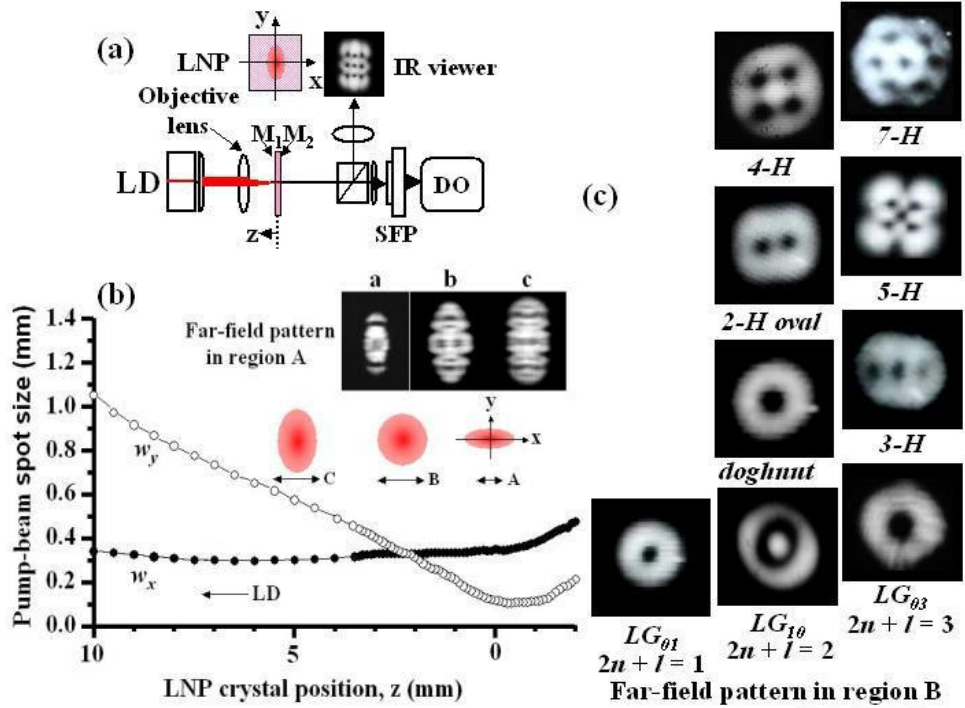

**Figure 1**: (a) Experimental setup. LD: laser diode, SFP: scanning Fabry-Perot interferometer, DO: digital oscilloscope. (b) LD pump beam profile as a function of LNP position. The pump-power region where each lasing pattern was observed is indicated by the arrows. Far-filed patterns observed in region A are depicted in the inset, where $w_x$ = 330 μm, $w_y$ = 70 μm, pump power $P$ = 293 mW (**a**), 366 mW (**b**) and 418 mW (**c**). Ordered transverse patterns were not observed outside the indicated regions. (c) Radial vortex lattice patterns on circular cylindrical coordinates observed with nearly symmetric large-aperture LD pumping. Pump power $P$; $2n + l = 1$ ($w_x$ = 330 μm, $w_y$ = 300 μm): 80 mW; $2n + l = 2$ ($w_x$ = 330 μm, $w_y$ = 330 μm): 105 mW (doughnut), 120 mW (oval), 140 mW (4 holes); $2n + l = 3$ ($w_x$ = 330 μm, $w_y$ = 360 μm): 160 mW (3 holes), 168 mW (5 holes), 175 mW (7 holes).

The observed lasing patterns can be categorized into three groups depending on the pump-beam profile, which featured complex patterns characterized by arrays of "dark" peaks rather than the grid of "bright" peaks of a conventional mode. Below, we show some example patterns.

A. *Optical billiard-like patterns consisting of non-orthogonal transverse modes*

Under the tight pump-beam focusing condition, i.e., sheet-pumping in the vicinity of z = 0, lasing patterns changed in a manner critically dependent on the crystal position (pump-beam shape). In this regime, orthogonal transverse modes were unable to enter oscillation, and lasing patterns were confirmed to consist of multiple non-orthogonal transverse modes with slightly different optical frequencies, as reported in ref. [8], which reported the observation of modal-interference-induced intensity modulations. A typical pump-dependent structural change of far-field lasing patterns observed in the present experiment is depicted in the inset of Fig. 1(b).

It should be noted that fundamental $TEM_{00}$ transverse mode operations were obtained independently of the pump power with a tightly focused circular LD pump beam, whose spot size was smaller than 100 μm, which was obtained by inserting an anamorphic prism between the LD and the microscope objective lens.

When the LNP position was gradually shifted from the focal point, a variety of lasing patterns appeared as the pump-beam size was increased (i.e., wide-aperture pumping) and the shape was changed.

B. *Radial vortex arrays*

When the pump-beam approached a symmetric shape (i.e., $w_x = w_y$), a doughnut-like Laguerre-Gaussian ($LG_{n,l}$) mode (radial index: $n = 0$, angular index: $l = 1$) on circular cylindrical coordinates was formed and a variety of patterns of radial vortex arrays appeared as the pump power was increased. The gain as well as refractive index confinement of lasing field depended on the pump power because the radial gain region and thermal refractive index increased (i.e., lens effect) with increasing pump power. Observed far-field patterns are shown in Fig. 1(c), where hole patterns are abbreviated as 4-H, 5-H, etc. in the figure. These patterns show surprisingly strong resemblance to patterns generated in an optically pumped $Na_2$ laser by Brambilla *et al.* [3]. They reproduced such lasing modes as doughnut, oval, and hole patterns by the coherent superposition of nearly degenerate modes (i.e., transverse mode locking in quadrature) belonging to the $2n + l = 1$ (two degenerate modes), 2 (three degenerate modes), and 3 (four degenerate modes) bands[4], where the resonant angular frequency is given by $\omega_{n,l} = (c/L)[q\pi + (2n + l + 1) \cos^{-1}\{\pm(g_1 g_2)^{1/2}\}]$. Here, $L$ is the cavity length, $q$ is the number of half-wavelengths along the cavity axis, and $g_j = 1 - L/R_j$ is the g-parameters of the resonator, where $R_j$ (j = 1, 2) is the radius of curvature of the mirror. For the platelet solid-state laser we used in the experiment, the g-parameters are predominantly determined by the pump-dependent thermal lens effect of the LNP crystal.

By measuring the far-field distribution at different distances from the laser, we observed that the complex patterns preserved the lateral functional structure against propagation. Thus, not only the total singularity charge propagation, but also the specific charge distributions are preserved in free space. Optical spectra measured with a scanning Fabry-Perot interferometer indicated single-frequency oscillations. This implies that observed complex patterns are formed through transverse locking of nearly degenerate modes belonging to the same band, i.e., coherent superposition of coexisting modes locked in phase and space quadrature, which ensures single-frequency operations. A similar structural change, resulting from the successive transverse locking among modes as the injection current was changed, was observed in a VCSEL [4].

C. *Rectangular vortex arrays*

As the crystal position was shifted further toward the LD, the LD pump-beam size increased and its shape became elongated along the y-axis, i.e., $w_x < w_y$. In this regime, a new class of vortex arrays was formed, starting from a higher-order Ince-Gaussian ($IG_{p,m}$) mode on elliptical coordinates [9], reflecting the asymmetric pump-beam shape.

Typical structural change of lasing patterns with increasing the pump power (i.e., lateral effective gain region) for different pump-beam shapes are shown in Fig. 2. In the case of (a), $IG^e_{3,1}$ mode with a relatively large ellipticity parameter ε, which resembles the corresponding Hermite-Gaussian mode, $HG_{1,2}$, appears at **a** and made a structural change to the 5-hole pattern as shown in column **b**. With increasing the pump power, mode-locking failed and $IG^e_{4,2}$ mode replaced the 5-hole (i.e, 5-H) pattern at **c**. Then, $IG^e_{4,2}$ mode made a structural change to the 8-hole (8-H) pattern as shown in column **d**. In the case of (b), 4-H (i.e., [2, 2] array) and 6-H ([2, 3] array) patterns appeared successively, although the lasing started from $IG^e_{3,1}$ mode similar to (a). As for (c), starting from the $IG^e_{4,2}$ mode with a smaller ε-value as compared with those in column **c** of (a) and (b), 6-H ([3, 2] array) and 9-H ([3, 3] array) patterns appeared successively. These ordered vortex array patterns maintained stably within the pump power change of ΔP ≈ 20 mW. From these experimental results, different IG-originated vortex patterns are considered to be formed depending on the pump (gain) profile and ellipticity parameter even if the starting IG mode possesses the same modal indices, *p* and *m*..

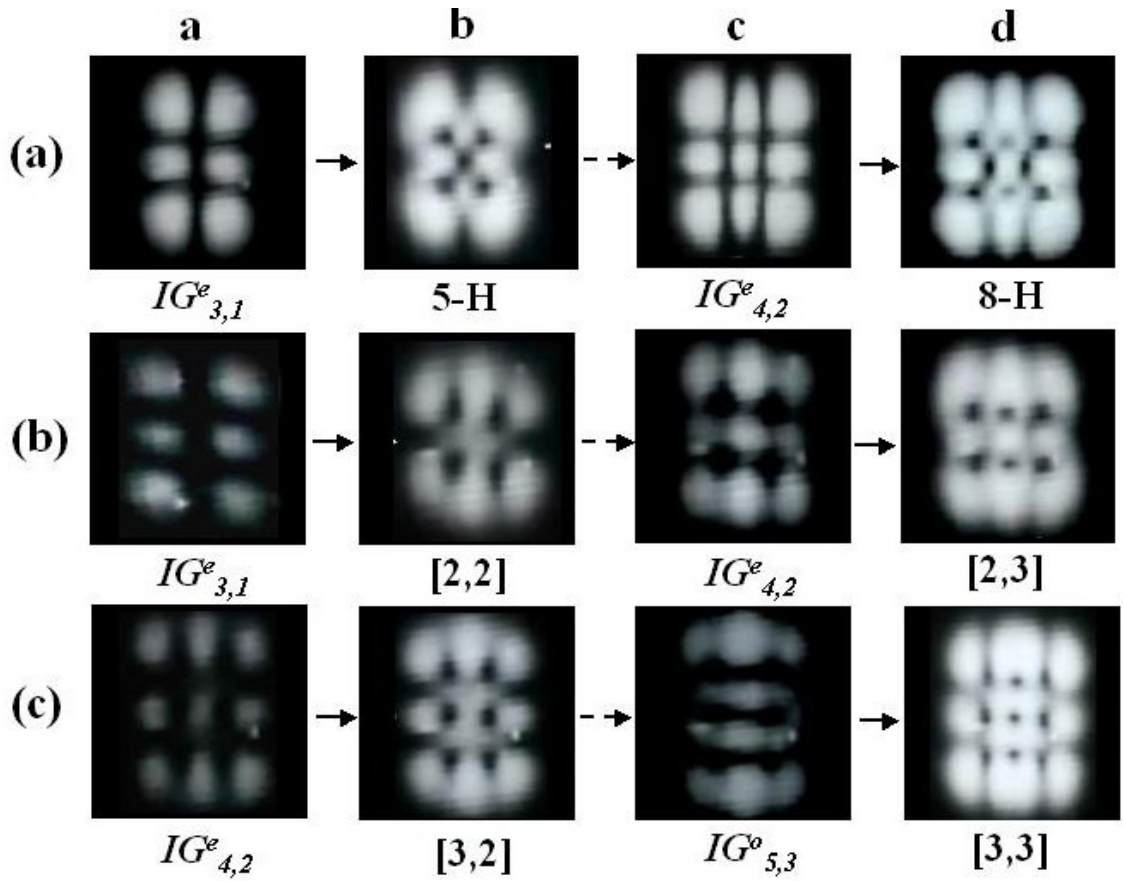

**Figure 2**: Pump-dependent structural change of vortex arrays with asymmetric large-aperture LD pumping. (a) $w_x$ = 300 μm, $w_y$ = 550 μm. column a: *P* = 177 mW; b: 191 mW; c: 195 mW; d: 198 mW. (b) $w_x$ = 300 μm, $w_y$ = 570 μm. column a: *P* = 164 mW; b: 190 mW; c: 200 mW; d: 207 mW. (c) $w_x$ = 300 μm, $w_y$ = 600 μm. column a: *P* = 156 mW; b: 164 mW; c: 195 mW; d: 201 mW.

All these vortex arrays have been theoretically reconstructed remarkably well by the locking of nearly degenerate IG modes in quadrature, as shown in Fig. 3, where the resonant angular frequency for $IG_{p,m}$ is given by replacing $(2n + l + 1)$ for LG modes by $(p + 1)$ [9]. Here, $i$ implies a phase shift of $\pi/2$. In all cases, modes to be locked to the first lasing IG modes, namely "paring modes" have less field overlapping, i.e., transverse cross-saturation of population inversions, and they tend to coexist with the first lasing modes with increasing the pump power.

The 5-H and 8-H patterns can be well reconstructed by frequency-degenerate even and odd IG modes, as shown in the first row of Fig. 3. In this case, there is no frequency competition among the modes, so cooperative mode locking [10] is necessary for the formation of stationary states only in the sense of phase locking with the relative phase among modes being fixed (3). For the [2, 2], [2, 3], [3, 2] and [3, 3] array patterns, on the other hand, nearly degenerate IG modes with the same parity must be locked to a common frequency in addition to the phase locking to establish cooperative mode locking leading to the vortex arrays shown in second and third columns of Fig. 3. Here, the common oscillation frequency falls within the range spanned by the mode-pulled frequencies of the coexisting modes through the third-order optical nonlinearity in lasers [10], as discussed later. The simple superposition of IG mode pairs in Figs. 3 without mode-locking resulted in totally different patterns. The present 4- and 5-hole patterns on the elliptic coordinates are different from those formed by the locking of more than two transverse modes [3] on the circular cylindrical coordinates shown in Fig. 2. The following points are obvious from the phase portraits, in which sold and open circles denote clock-wise and anti-clock-wise phase rotations, respectively: (1) vortices in the same row or column are rotating in same direction, but the "starting" phases of nearest neighbours differ by $\pi$ and (2) vortices in the neighbouring row or column are rotating in the opposite direction.

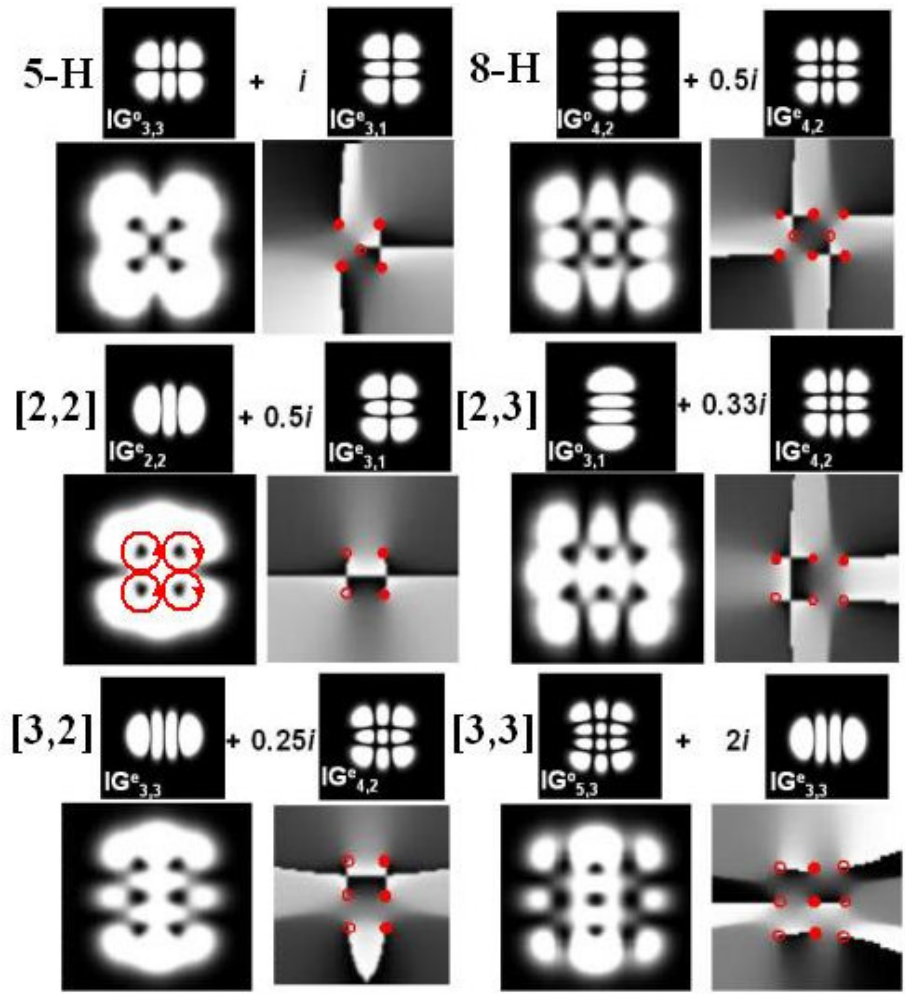

**Figure 3**: Theoretically reconstructed vortex arrays. Used ellipticity parameters: $\varepsilon = 50$ for 8-H array, $\varepsilon = 30$ for 5-H, [2, 3] arrays, and $\varepsilon = 10$ for [3, 2], [3, 3] arrays. Sold circles: clockwise phase rotation, open circles: anti-clockwise phase rotation. The starting phases of nearest neighbour vortices in the same row or column are shifted by $\pi$.

The coherent superposition of degenerate even and odd IG modes, i.e., $HIG_{p,m} = IG^e_{p,m} \pm i\, IG^o_{p,m}$, has been proposed by Bandres and Gutierrez-Vega for generating helical Ince-Gaussian (HIG) beams [9], and elliptic optical vortices with in-line phase singularities inside have been generated with a liquid crystal display (i.e., spatial light modulator) [11]. In our experiment, the coherent superposition (i.e., cooperative locking) of nearly degenerate IG modes, which can lase simultaneously through transverse spatial hole-burning, was established by the intrinsic laser nonlinearity. We have observed various other vortex array patterns besides those shown in Fig. 2, depending on the LD-beam focusing conditions.

When the pump power was increased further, vortex array patterns on elliptical coordinates gradually changed their shape to rectangular-type arrays on Cartesian coordinates, with the number of vortices, i.e., $q \times r$, increasing successively due to the increase in the gain region.

An example of pump-dependent structural changes is shown in Fig. 4. A [3, 3] single-frequency rectangular vortex array was formed in **a**. With increasing pump power, the rectangular vortex array, which consisted of 5- and 4-hole fundamental cell structures in Figs. 2(a) and 2(b) indicated by dashed boxes, appeared in **b** in the transition process to the single-frequency [6, 4] vortex array shown in **c**. In the high pump-power region, on the other hand, such a lasing rectangular vortex-array pattern was spontaneously clustered into multiple vortex sub-lattice structures differentiated by "dark lines", as indicated by the white arrows in **d**, **e** and **f**. At this moment, the LNP laser was found to exhibit multiple-frequency operations as shown in optical spectra, in which each lattice group had its own frequency. Similar multiple-frequency vortices crystals were observed in the high injection-current regime of VCSELs [4]. Magnified views of central regions are also shown in Fig. 4.

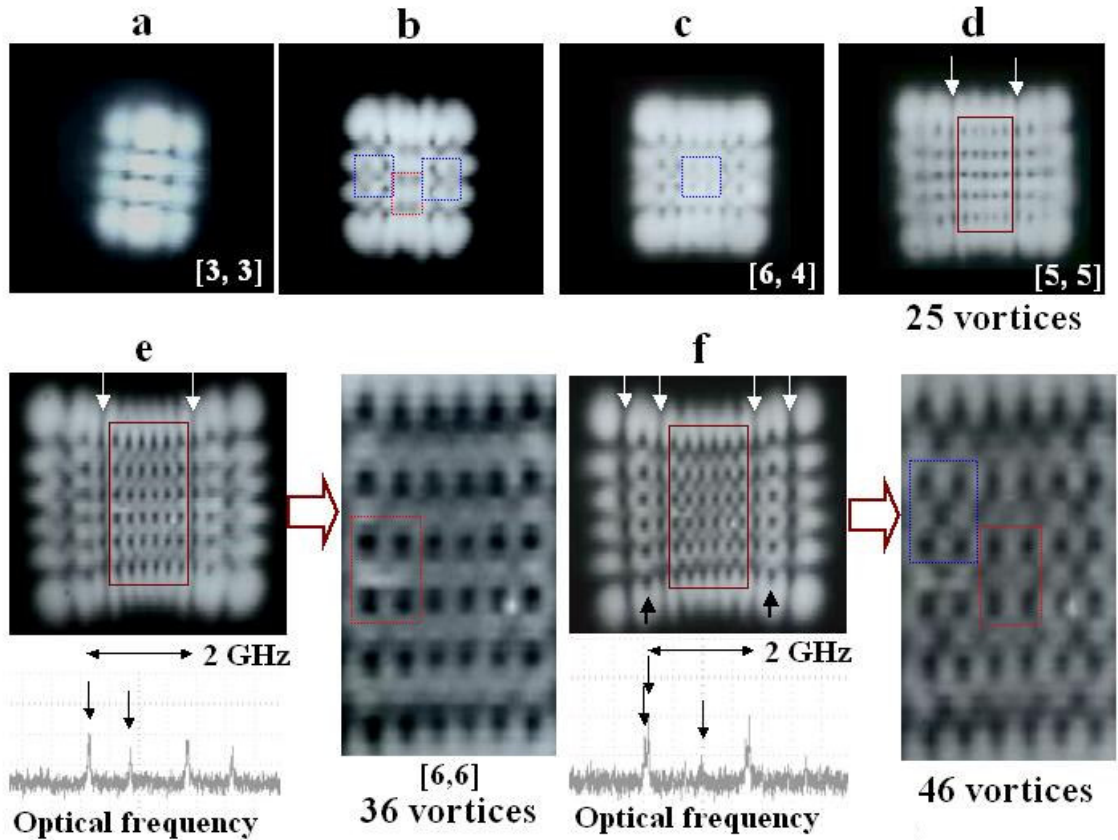

**Figure 4**: Pump-dependent structural change of [*q*, *r*] vortex array and magnified views observed in the high pump-power region. $w_x$ = 300 µm, $w_y$ = 620 µm. a: *P* = 233 mW; b: 267 mW; c: 284 mW; d: 430 mW; e: 498 mW; f: 515 mW.

Closely packed [6, 6] vortex crystal with 36 vortices was formed in an area of about 500 µm$^2$ in **e**. Here, the fundamental 4-hole unit cell structure shown in Fig. 2(b) is seen clearly, as indicated by the dashed box. When the pump power was increased to **f**, the 5-hole unit cell structure shown in Fig. 2(a) appeared together with the 4-hole unit cell structure forming 46 vortices crystal in the central region, where one-dimensional vortex chains were formed symmetrically outside the central vortex crystal, as indicated by black arrows.

The key mechanism of transverse locking in thin-slice solid-state lasers is attributed to the saturation-type of third-order (i.e., $\chi^{(3)}$) optical nonlinearity in thin-slice solid-state lasers similar to VCSELs *(4)*. In thin-slice solid-state lasers, the longitudinal mode spacing determined by $\Delta\lambda = \lambda^2/2nL_c$ ($\lambda$: lasing wavelength, $n$: refractive index, $L_c$: crystal thickness) is comparable to the gain bandwidth, $\Delta\lambda_g$. (As for the present LNP laser, $\Delta\lambda = 1.14$ nm where $\Delta\lambda_g = 1.7$ nm). is comparable to the gain bandwidth. Consequently, the oscillation frequency of the first lasing longitudinal mode is detuned from the gain center in general. In such detuned lasers, a change in the intensity-dependent refractive index (real part of the electric susceptibility $Re(\chi^{(3)})$) is expected through gain ($Im(\chi^{(3)})$) saturation [10]. In LDs, such a third-order optical nonlinearity is expressed by the $\alpha$ parameter, e.g., $\alpha = Re(\chi^{(3)})/\varepsilon_0 n_0 = 3$–$6$ for LDs ($\varepsilon_0$: vacuum dielectric constant, $n_0$: linear refractive index). Similar $\chi^{(3)}$ nonlinearity exists in thin-slice solid-state lasers, i.e., $n = n_0 + n_2 I$ (*I*: light intensity); $n_2 = g_0\alpha(\lambda/4\pi\, I_s)$ ($g_0$: small signal gain, $I_s$: saturation intensity), where $\alpha = 1$–$2$ for LNP lasers [12]. The resultant phase-sensitive interaction among nearly degenerate transverse modes results in frequency-locking with the fixed relative phase depending on the pump power, i.e., modal intensities.

In summary, we have demonstrated spontaneous formations of a variety of lasing patterns in a thin-slice, wide-aperture solid-state laser under different LD pump-beam profiles and sizes. The radial-symmetric vortex lattices reported in this paper had patterns that showed surprisingly strong resemblance to patterns generated in other lasers, e.g., $Na_2$ laser and VCSELs [3,4]. On the other hand, for asymmetric pump-beam profiles with increased pump area, [$q$, $r$]-type vortex array beams were found to be generated spontaneously, originating from Ince-Gaussian modes on the elliptic coordinates, in which locking IG pairing modes are automatically determined through transverse spatial hole-burning. These spontaneously generated laser array beams consisting of a large number (e.g., 36 and 46) of ordered optical vortices, which can be controlled merely by the pump-beam focusing condition without using a well-designed phase plate (e.g., spatial light modulator and hologram), can be used to construct a new class of optical tweezers [13] and atom traps in the form of two-dimensional arrays as well to study the transfer of angular momentum to microparticles or atoms (Bose-Einstein condensate) [6]. Growing desired vortex crystals could be possible by adding a tightly focused off-axis [14] or azimuthal pump beam [15] to the main wide-aperture pump beam for lasing a desired seed IG mode from which vortex arrays are formed with increasing the pump power of the main beam.